\documentclass{IEEEtran}
\author{Ekaterina Bayguzina and Bruno Clerckx
\thanks{The authors are with the Department of Electrical and Electronic Engineering,
Imperial College London, London SW7 2AZ, U.K. (e-mail: ekaterina.bayguzina08@imperial.ac.uk; b.clerckx@imperial.ac.uk).}
\thanks{
This paper was presented in part at the IEEE 19th International Workshop on Signal Processing Advances in Wireless Communications, Kalamata, Greece, June 2018 \cite{SPAWC_paper}. 
This work was supported in part by the EPSRC of the UK, under grants EP/K502856/1, EP/L504786/1 and EP/P003885/1.}

\thanks{Manuscript version 20 August 2019.}}

\title{Asymmetric Modulation Design for Wireless Information and Power Transfer with Nonlinear Energy Harvesting}
\date{}

 \usepackage{graphicx}
 \usepackage{capt-of}
\usepackage{amsmath}
 \usepackage{amssymb}
 \usepackage[hidelinks=true]{hyperref}
 \usepackage{breqn}

 \DeclareMathOperator*{\argmax}{arg\,max}

\begin{document}
\begin{titlepage}
$\copyright$ 2019 IEEE. Personal use of this material is permitted. Permission from IEEE must be obtained for all other uses, in any current or future media, including reprinting/republishing this material for advertising or promotional purposes, creating new collective works, for resale or redistribution to servers or lists, or reuse of any copyrighted component of this work in other works.
\end{titlepage}
\maketitle
\begin{abstract}
Far-field wireless power transfer (WPT) and simultaneous wireless information and power transfer (SWIPT) have become increasingly important in radio frequency (RF) and communication communities recently. The problem of modulation design for SWIPT has however been scarcely addressed. 
In this paper, a modulation scheme based on asymmetric phase-shift keying (PSK) is considered, which improves the SWIPT rate-energy trade-off region significantly. The nonlinear rectifier model, which accurately models the energy harvester, is adopted for evaluating the output direct current (DC) power at the 
receiver. The harvested DC power is maximized under an average power constraint at the transmitter and a constraint on the rate of information transmitted via a multi-carrier 
signal over a flat fading channel. As a consequence of the rectifier nonlinearity, this work highlights that asymmetric PSK modulation provides benefits over conventional symmetric PSK modulation in SWIPT and opens the door to systematic modulation design tailored for SWIPT. 
\end{abstract}
\section{Introduction}
Radio frequency (RF) wireless power transfer (WPT) has become more important in recent years with the advent of mainly battery free wireless sensor networks (WSN) 
and Internet of Things (IoT). At the same time, new wireless communications systems enabling simultaneous wireless information and power transfer (SWIPT) are widely investigated in the research community \cite{BClerckx_2019}. For the purpose of SWIPT, RF-based WPT is particularly relevant, due to the common use of RF signals in communications, as well as the omnipresence of various radioelectric waves in the ambient environment \cite{Visser}, \cite{Pinuela}. Moreover, WPT via radio waves is able to cover longer range as opposed to near-field WPT via inductive coupling or magnetic resonant coupling \cite{Shinohara}, \cite{Valenta}. 
With the unification of wireless power and wireless information transfer (WIT), many opportunities arise, such as ubiquitous power
accessibility,  ``green'' self-sustainable operation, battery free (or reduced-size) lower cost devices, and true mobility with no wires required to deploy a network \cite{Toward1G}.
The integration of wireless power and wireless communications brings new challenges and opportunities, and calls for a paradigm shift in wireless network design. As a result, numerous new research problems need to be addressed that cover a wide range of disciplines including communication
theory, information theory, circuit theory, RF design, signal processing, protocol design, optimization, prototyping, and experimentation. 

\par Of particular interest to any communication system is the notion of modulation. Efficient modulations have been designed for several decades to maximize spectral efficiencies and minimize error probability of communication systems \cite{fundamentals}. However, with the emerging SWIPT communication paradigm, modulation needs to be re-designed and re-thought. Indeed, while modulation affects the spectral efficiency of the information transfer, it also influences the amount of DC power that can be harvested. This motivates the central question of this work: How to design efficient modulation for SWIPT?

\par The effect of various conventional communication modulation schemes on the energy conversion efficiency (ECE) as compared to continuous wave (CW) transmission has been considered by the RF community \cite{FSKrectenna}--\cite{RF_harvesting_DM}. In particular, the ECE has been shown to be dependant on various conditions, such as modulation type, input power, and rectifier circuit parameters. 
In \cite{FSKrectenna}, FSK modulated signals were shown to be detrimental in terms of harvested power due to the mixing operation of the diode-based rectifier. In \cite{Sakaki_2014}, phase and amplitude variations introduced by QPSK and 16QAM were demonstrated to decrease the ECE as compared to the CW. However, in \cite{Mod_Scheme_RF_DC_Japan}, it was established that although the energy harvester (EH) performance is degraded with 
symmetric modulated signals such as QPSK and 16QAM 
at high input powers, these modulation schemes improve the ECE at low input powers significantly. 
In \cite{RF_harvesting_DM}, it was shown that the ECE, achieved via multitone and modulated signals, depends on the circuit optimization, mainly on the matching network and the load. 
Accordingly, a signal with a time-varying envelope can enhance the ECE for a certain range of load values and input power levels in relation to a CW signal with the same average power. 
\par However, modulation schemes designed specifically for the purpose of SWIPT have scarcely been addressed 
\cite{SWIPT_PAPR}--\cite{Zewde_Gursoy}. 
In particular, 
modulation for SWIPT was performed by varying the peak-to-average-power ratio (PAPR) of a multisine signal in \cite{SWIPT_PAPR}; biased ASK was proposed in \cite{Claessens}; and the symbol probabilities of finite rectangular/square constellations were optimized in \cite{Zewde_Gursoy}. 
Moreover, most of the existing results in the diverse literature on SWIPT are based on the linear model of the EH \cite{Zewde_Gursoy}--\cite{Rajashekar}. Yet, the linear EH model is limited in terms of accuracy, and efforts have recently been devoted to WPT and SWIPT design under nonlinear EH modelling \cite{Clerckx:2016b}--\cite{Alevizos}. Remarkably, systematic waveform design exploiting the EH nonlinearity allows to achieve a higher ECE (and overall end-to-end power transfer efficiency) in a practical rectenna circuit with the use of multiple sinewaves \cite{Clerckx:2016b}, \cite{letter_paper}--\cite{Huang}. 
\par Leveraging \cite{Clerckx:2016b}, SWIPT signal design accounting for the rectifier nonlinearity was studied in \cite{Clerckx:SWIPT_long}.  It was concluded in \cite{Clerckx:SWIPT_long} that the rectifier nonlinearity radically changes the design of SWIPT. Indeed ``it favours a different waveform, modulation, input distribution and transceiver architecture as well as a different use of the RF spectrum. Exploiting the rectifier nonlinearity in the SWIPT design also makes a more efficient use of the resources by enabling enlarged rate-energy regions compared to those obtained by ignoring the nonlinearity in the system design.''
Motivated by those observations, the optimal input distribution of SWIPT subject to nonlinear power constraints was studied in \cite{Morteza1}, \cite{Morteza2}. Remarkably, it was shown that the capacity of an additive white Gaussian noise (AWGN) channel under transmit average power and nonlinear delivered power constraints is the same as the capacity of an AWGN channel under an average power constraint. In other words, the capacity of an AWGN channel is independent of the value of the delivered power constraint. The capacity 
can be arbitrarily approached by using time sharing between distributions with high amount of information, e.g. circularly symmetric complex Gaussian (CSCG) inputs, and distributions with high amount of power reminiscent of flash signaling exhibiting a low probability of high amplitude signals.

\par In this paper, we investigate modulation for SWIPT considering the accurate nonlinear model for the EH at the receiver and flat fading environment. We study modulation of information symbols onto the multi-carrier energy-carrying waveform, resulting in a unified SWIPT waveform. Uniquely, we introduce asymmetric PSK modulation and show its suitability for SWIPT. In the following, we summarize the contributions of this paper.
\begin{itemize}
\item The nonlinear model of the rectifier is extended in order to account for the randomness in phases of carriers. In particular, we choose the phases to be independent and identically distributed (i.i.d.) random variables with uniform distribution. 
We show through analytical derivation that the scaling of the harvested energy with the number of carriers exhibits increasing losses as the phase range increases, degrading the benefit of using a multi-carrier signal. However, departing from the in-phase condition for maximum power transfer opens up a possibility of transmitting information with symbols distributed uniformly over a non-zero phase range.
\item We formulate the asymmetric $M$PSK modulation scheme and obtain a metric for the harvested energy given 
mass probabilities of the input symbols. We then derive the achievable information rate for a flat fading channel, and consequently, obtain the optimal input distribution for the proposed modulation scheme maximizing the rate only. Finally, we characterize the SWIPT rate-energy region considering the designed modulation and optimize the probabilities of the input symbols under rate and energy constraints. 
\item We conclude the results by providing the complimentary numerical results. We support all the observations with simulations of a practical rectenna circuit. 
The main observations are that 1) the nonlinear EH model allows to optimize the probabilities of symbols for symmetric and asymmetric PSK modulations, enlarging the rate-energy region; 2) asymmetric PSK is advantageous over symmetric PSK in terms of harvested energy, and additionally, it is not too detrimental in terms of information rate, as such phase range can be obtained for a given signal-to-noise ratio (SNR) in order to balance the rate-energy maximization mechanisms; 3) analysis of optimized probabilities of symbols confirms that maximum power transfer is achieved with in-phase multi-carrier transmission, whereas maximum rate corresponds to the optimal input distribution obtained via the Blahut-Arimoto algorithm.
\end{itemize}

\par \textit{Organization:} In Section 
II, the SWIPT system model is introduced. In Section 
III, the design of the asymmetric PSK modulation and the optimization of the rate-energy region is studied. In Section 
IV, the simulation results are presented. Finally, Section 
V concludes the paper. 
\par \textit{Notation:} Throughout this paper, the operators $\mathcal{E}\{\cdot\}$ and $\mathbb{E}\{\cdot \}$ refer to the average over time and the expectation over statistical randomness, respectively. Upper case letters stand for random variables and random processes, except for in relation to circuit notations. The probability density (mass) function of a continuous (discrete) random variable $X$ is denoted by $p_X(x)$, and the probability of an event is denoted by $\mathrm{Pr}(\cdot )$. 
$\Re\{\cdot\}$ and $\Im\{\cdot \}$ indicate real and imaginary operators, respectively. The standard circularly symmetric complex Gaussian distribution is denoted by $\mathcal{CN}(0,1)$. $Var\{\cdot\}$ refers to the variance of a random variable. $U[-a,a]$ denotes the uniform distribution over the interval $[-a,a]$. The convolution operator is denoted by $\ast$.

\section{SWIPT System Model}\label{system_model}
\begin{figure}
\centering
\includegraphics[width=\linewidth]{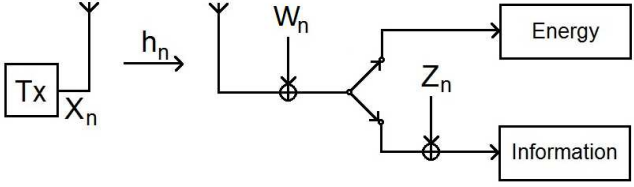}
  \caption{System model with the receiver simultaneously decoding information and harvesting energy.}
  \label{joint_EH_ID}
\end{figure}

We study the system model illustrated in Fig. \ref{joint_EH_ID}, in which a single-input single-output (SISO) point-to-point SWIPT is considered and the receiver is assumed to be able to simultaneously decode information and harvest energy. We consider a multi-carrier/band transmission (with single-carrier being a special case) consisting of $N$ orthogonal subbands where the $n$th subband has carrier frequency $f_n$ and equal bandwidth $B_s$, $n=0,\ldots,N-1$. The carrier frequencies are evenly spaced such that $f_n=f_0+n\Delta_f$ with $\Delta_f$ denoting the inter-carrier frequency spacing (with $B_s \leq \Delta_f$). The channel, over which the signal is transmitted, is assumed to be flat fading (when $(N-1)\Delta_f$ is much smaller than the channel coherence bandwidth) and is subject to AWGN, originating from the receiver antenna and the RF-to-baseband processing at the information decoder. Throughout the paper, we assume that the power of the processing noise $Z_n$ is much larger than the power of the antenna noise $W_n$ and that the effect of the antenna noise is negligible for energy harvesting purposes \cite{BClerckx_2019}.

\subsection{Transceiver Design}
Considering a multi-carrier transmission, the channel input is given as
\begin{align}\label{WPT_waveform}
x(t)=\Re\left\{\sum_{n=0}^{N-1}X_{n}e^{j2\pi f_n t}\right\}
\end{align}
with $X_{n}=S_ne^{j\Phi_{n}}$, where $S_n$ and $\Phi_{n}$ refer to the amplitude and phase of the $n^{th}$ carrier at frequency $f_n$. 
The transmitter is subject to a transmit power constraint $\mathbb{E}\left\lbrace\left|X_n\right|^2\right\rbrace
\leq P$.

\par The transmitted signal $x(t)$ propagates through a multipath channel. The received signal at the receiver is modelled as
\begin{equation}
y(t)=\Re\left\{\sum_{n=0}^{N-1} h_n X_n e^{j 2\pi f_n t}\right\},
\label{received_signal_ant_m}
\end{equation}
where $h_{n}$ is the channel frequency response at frequency $f_n$. 

\subsection{Antenna Equivalent Circuit Model}
\begin{figure}
\centering
\includegraphics[width=\linewidth]{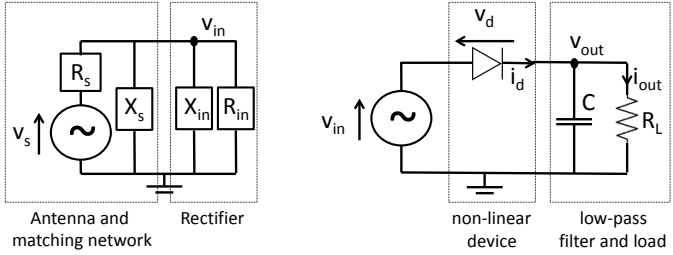}
  \caption{Antenna equivalent circuit (left) and a single series diode rectifier (right).}
  \label{antenna_model}
 \end{figure}

The antenna model reflects the power transfer from the
antenna to the rectifier through the matching network. As illustrated in Fig. \ref{antenna_model} (left), a lossless antenna
can be modelled as a voltage source $v_s(t)$ followed by a
series resistance $R_{s}$ and a parallel reactance $X_s$. The rectifier is modelled as a resistance $R_{in}$ in parallel with a reactance $X_{in}$. Assuming perfect matching ($R_{in} = R_{s}$, $X_{in} = -X_s$), all the available RF power $P_{in,av}$ is transferred to the rectifier and absorbed by $R_{in}$, so that $P_{in,av} = \mathbb{E}\left\lbrace\mathcal{E}\left\lbrace\left|v_{in}(t)\right|^2\right\rbrace\right\rbrace/R_{s}$. Since $P_{in,av} =\mathbb{E}\left\lbrace\mathcal{E}\left\lbrace\left|y(t)\right|^2\right\rbrace\right\rbrace$, $v_{in}(t)=y(t)\sqrt{R_{s}}$ and $v_s(t)$ can be formed as
\begin{equation}\label{vs_eq}
v_s(t)=2y(t)\sqrt{R_{s}}.
\end{equation}

\subsection{Rectifier and Diode Nonlinear Model}
Consider a rectifier composed of a single series diode followed by a low-pass filter with a load as in Fig. \ref{antenna_model} (right). Denote the voltage drop across the diode as $v_d(t)=v_{in}(t)-v_{out}(t)$ where $v_{in}(t)$ is the input voltage to the diode and $v_{out}(t)$ is the output voltage across the load resistor. A tractable behavioural diode model is obtained by Taylor series expansion of the diode characteristic equation $i_d(t)=i_s \big(e^{\frac{v_d(t)}{n v_t}}-1 \big)$ (with $i_s$ the reverse bias saturation current, $v_t$ the thermal voltage, and $n$ the diode ideality factor) around a quiescent operating point $v_d=a$, namely
\begin{equation}\label{DiodeCurrent}
i_d(t)=\sum_{i=0}^{\infty }k_i \left(v_d(t)-a\right)^i,
\end{equation}
where $k_0=i_s\big(e^{\frac{a}{n v_t}}-1\big)$ and $k_i=i_s\frac{e^{\frac{a}{n v_t}}}{i!\left(n v_t\right)^i}$, $i=1,\ldots,\infty$ \cite{Clerckx:2016b}.

\par Assume a steady-state response and an ideal low pass filter, such that $v_{out}(t)$ is at a constant DC level. Choosing $a=\mathbb{E}\left\{ \mathcal{E} \left\{ v_d(t) \right\} \right\}=-v_{out}$, \eqref{DiodeCurrent} can be simplified as
\begin{equation}\label{polynomialSeries}
i_d(t)=\sum_{i=0}^{\infty }k_i v_{in}(t)^i=\sum_{i=0}^{\infty }k_i R_s^{i/2} y(t)^i.
\end{equation}

\par In order to make the model and the modulation design more tractable, we truncate 
\eqref{polynomialSeries} to the fourth order that is sufficient to retain the fundamental nonlinear behaviour of the diode \cite{Clerckx:2016b}, \cite{Boaventura:2011}.
After truncation and taking the average over time and the expectation over statistical randomness, the DC component of $i_{d}(t)$ 
is approximated as $i_{out}\approx k_0+z_{DC}$ where 
\begin{equation}
z_{DC}=k_2 R_{s}\mathbb{E}\left\{ \mathcal{E}\left\{y(t)^2\right\} \right\}+k_4 R_{s}^2\mathbb{E}\left\{ \mathcal{E}\left\{y(t)^4\right\} \right\}.
\label{zdc}
\end{equation}
In \eqref{zdc}, the coefficients $k_2=0.0034$ and $k_4=0.3829$ are calculated for a typical zero-bias Schottky diode with $i_s=5 \mu A$ and $n=1.05$, and $v_t=25.86 mV$. This model has been validated for
the design of multisine waveform in 
\cite{Clerckx:2016b}, \cite{letter_paper}, \cite{Clerckx_TD_2018} using
circuit simulators with various rectifier topologies and input
power and in \cite{Clerckx_TD_2018,Junhoon_prototyping} through prototyping and experimentation. It has also been validated using
circuit simulator for modulated single-carrier \cite{Morteza2} and modulated multi-carrier waveform with Gaussian input \cite{Clerckx:SWIPT_long}.

\section{Modulation Design for SWIPT}\label{section_modulation}
\par Considering a flat fading channel with gains given as $h_n=1$ for $n=0,\ldots,N-1$, in the following, we first study the scaling of the harvested power at the receiver with respect to the number of carriers in Section III.A, as well as the variation in the phase range of the transmitted signal. This analysis demonstrates 
how the random phase affects the harvested power by reducing the gains of the multi-carrier transmission. Moreover it gives an insight into how the adverse effects of the phase variation can be mitigated by efficient modulation design and why SWIPT modulations differ from conventional communication modulations. Next, in Section III.B, we consider the $M$PSK modulation design for SWIPT purposes. In Section III.C, we study the ultimate information-theoretic achievable rate for the considered $M$PSK modulation. Finally, in Section III.D, we consider optimization of the harvested power under an average power constraint and a received information rate constraint.

\subsection{Scaling Law for Uniformly Distributed Random Phase}\label{section_scaling_uniform}
An analytical expression for $z_{DC}$ in \eqref{zdc} is now derived using the multi-carrier signal with uniform power allocation across the carriers with $S_n=s=\sqrt{\frac{2P}{N}}$, $n=0,\ldots,N-1$ and uniformly distributed random phases. Choosing the phases of different carrier frequencies $f_n$, $n=0,\ldots,N-1$ to be i.i.d. with $\Phi_{n} \sim U[-\delta,\delta]$, $n=0,\ldots,N-1$,  $\delta \leq \pi$,
for a flat fading channel, 
the received signal $y(t)$ is given by
\begin{equation}
y(t)=\sqrt{\frac{2P}{N}}\sum_{n=0}^{N-1}\cos(2\pi f_n t+\Phi_n).
\label{TxSignalM1}
\end{equation}
The second and fourth order terms in $z_{DC}$ are further simplified as follows. By taking the time average and expectation over the randomness of the signal, we have:

\begin{equation}
\mathbb{E}\left\{ \mathcal{E}\left\{ y(t)^2 \right\}\right\}= P,
\label{2ndOrderSumVaryingPhase}
\end{equation}
\begin{multline}
\mathbb{E}\left\{ \mathcal{E}\left\{y(t)^4\right\}\right\}
=\frac{3P^2}{2N^2}\sum_{\substack{n_0,n_1,n_2,n_3 \\ n_0+n_1=n_2+n_3 }}\mathbb{E}\left\{\cos\left(\Phi_{n_0} \right. \right.\\
\left. \left. +\Phi_{n_1}-\Phi_{n_2}-\Phi_{n_3}\right)\right\}.
\label{4thOrderSumVaryingPhase}
\end{multline}
Note that in \eqref{2ndOrderSumVaryingPhase} and \eqref{4thOrderSumVaryingPhase}, we have neglected the noise power. Else, we assume that the average power constraint at the transmitter is satisfied with equality. For details on the derivation of \eqref{2ndOrderSumVaryingPhase} and \eqref{4thOrderSumVaryingPhase}, refer to Section III.C in \cite{Clerckx:2016b}\footnote{Equations \eqref{2ndOrderSumVaryingPhase} and \eqref{4thOrderSumVaryingPhase} can be obtained, respectively, from equations (12) and (14) of \cite{Clerckx:2016b} by substituting in the latter $A_n=h_n=1$, $s_n=s=\sqrt{\frac{2P}{N}}$ and $\psi_n = \Phi_n$ for $n=0,\ldots,N-1$, for a single transmit antenna.}.

\par
\begin{figure}
\centering
\includegraphics[width=\linewidth]{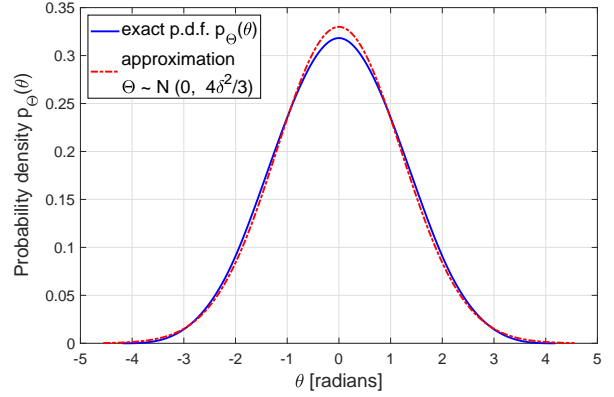}
\caption{The exact and normal approximation of the p.d.f. of $\Theta$ with $\delta=\pi/3$.}
\label{distributions_overlay}
\end{figure}

 Denoting the random term in (\ref{4thOrderSumVaryingPhase}) as
\begin{equation}
\Theta=\Phi_{n_0}+\Phi_{n_1}-\Phi_{n_2}-\Phi_{n_3},
\label{Theta_eq}
\end{equation}
the resultant probability density function (p.d.f.) of $\Theta$ is obtained by convolving four uniform distributions $\Phi_{{n}_i} \sim U [-\delta,\delta]$, $i=0,\ldots,3$. The exact distibution of $\Theta$ can also be approximated with a normal distribution. In \cite{uniform_to_normal}, it is shown that the sum of $I$ i.i.d. and uniformly distributed random variables converges to the normal distribution extremely fast. Accordingly, we approximate the distribution of $\Theta$ in \eqref{Theta_eq} by a normal distribution $\mathcal{N}(\sum_{i=0}^{3} \mathbb{E}\{\Phi_{n_i}\},\sum_{i=0}^{3} {Var}\{\Phi_{n_i}\})$ with
\begin{equation}
\mathbb{E}\{\Phi_{n_i}\}=0,
\end{equation}
\begin{equation}
{Var}\{\Phi_{n_i}\}=\frac{\delta^2}{3}.
\end{equation}
The distribution of $\Theta$ is thus approximated as $\Theta \sim \mathcal{N}(0,\frac{4\delta^2}{3})$. The difference between the exact distribution and the normal approximation for $\delta=\pi/3$ is shown in Fig. \ref{distributions_overlay}.

\par Next, the expected mean of the random variable $\cos(\Theta)$ is obtained as follows:
\begin{equation}
\mathbb{E}\{\cos(\Theta)\}=\Re\left\{\mathbb{E}\{e^{j\Theta}\}\right\}=e^{\frac{-2\delta^2}{3}}.
\label{cos_theta}
\end{equation}

\par Finally, noting that there are $N(2N^2+1)/3$ terms in the sum of \eqref{4thOrderSumVaryingPhase}, the scaling law for $z_{DC}$ for the signal $y(t)$ in \eqref{TxSignalM1} is obtained as 
\begin{equation}
z_{DC}(\delta)\simeq k_2 R_sP+k_4 R_s^2\frac{(2N^2+1)}{2N}P^2e^{\frac{-2\delta^2}{3}}.
\label{Scaling_law_phase}
\end{equation}

It is observed that the harvested power given by the $z_{DC}$ metric  increases with $N$
but decreases with $\delta$. The rate of scaling of $z_{DC}$ with $N$ is diminished with enlarging the phase range, until it becomes nearly flat for $\Phi_{n} \sim U[-\pi,\pi]$. Moreover, the case of $\delta=0$ is equivalent to the scaling exhibited by the deterministic multisine signal in a frequency-flat channel in \cite{Clerckx:2016b}, whereas the case of $\delta=\pi$ leads to a scaling behaviour similar to the multi-carrier signal modulated with CSCG inputs in \cite{Clerckx:SWIPT_long}. Thus, departing from the in-phase condition for maximum power transfer 
makes it possible 
to simultaneously transfer information with a phase-modulated multi-carrier signal.

\textit{Remark 1:} While \eqref{Scaling_law_phase} is obtained as a result of the truncation of \eqref{polynomialSeries} to the fourth order and the Gaussian approximation of the distribution of $\Theta$ in \eqref{Theta_eq}, the observations above still hold if these approximations are removed. Firstly, considering as an example a typical input power of $-20$dBm at the EH receiver with the diode parameters from Section II.C and $N=8$, the ratio of the second order term to the fourth order term in \eqref{Scaling_law_phase} can be obtained as  $\approx \frac{k_2 e^{\frac{2\delta^2}{3}}}{k_4 R_s  N P }= 2.2e^{\frac{2\delta^2}{3}}$. Therefore, the second order term is dominating over the fourth 
(with the contribution of the fourth order term decreasing with larger $\delta$), and truncation to the fourth order is sufficient to accurately capture the diode nonlinear behaviour in the low power regime. Secondly, the speed of the convergence of the result of the convolution of the identical uniform distributions of $\Phi_{n_i}$ in the argument of $\cos( \cdot)$ to the normal distribution is extremely fast \cite{uniform_to_normal} and increasing with the number of $\Phi_{n_i}$. 
Therefore, if \eqref{polynomialSeries} is truncated to a higher order, the difference due to the Gaussian approximation can be attributed at most to the approximation of the p.d.f. $p_{\Theta}(\theta)$ in the fourth order term (which can be considered a minimal difference, as shown in Fig. \ref{distributions_overlay}) rather than in the higher order terms.

\subsection{Asymmetric $M$PSK Modulation Design for SWIPT}
\par The scaling law for $z_{DC}$ \eqref{Scaling_law_phase} motivates the use of modulation schemes, in which symbols are distributed in a limited phase range. Accordingly, this  allows to obtain higher gains in harvested power.

\begin{figure}
\centering
\includegraphics[width=0.9\linewidth]{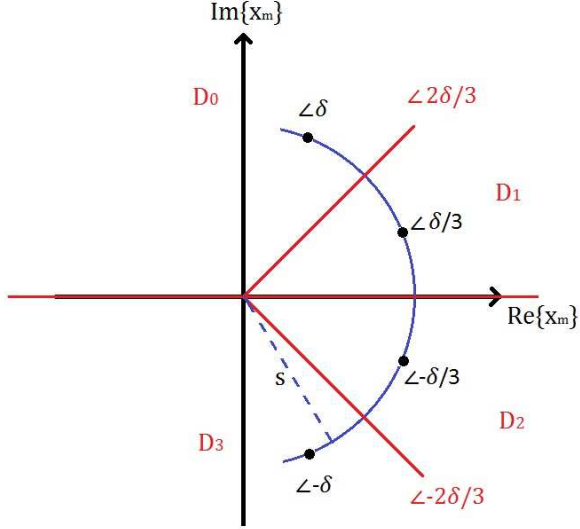}
\caption{Asymmetric 4PSK constellation with $\delta=\pi/3$. MAP decision boundaries are shown by red lines.}
\label{MPSK_figure}
\end{figure}

Motivated by the WPT observations in Section III.A
, for asymmetric $M$PSK modulation, the symbols are given as 
\begin{multline}
X_n\in \bigg\{x_m=se^{\pm j (2m+1)\delta/(M-1)},\ m=0,1,\ldots ,\frac{M}{2}-1\ : \\
|\delta|<\pi\bigg\},\ n=0,\ldots,N-1.
\label{mpsk} 
\end{multline}
Note that choosing $\delta=\pi$ yields the standard symmetric $M$PSK modulation:
\begin{multline}
X_n\in \bigg\{x_m=s e^{ \pm j(m+1)\pi/M},\ m=0,1,...\ ,\frac{M}{2}-1\ : \\
|\delta|=\pi\bigg\},\ n=0,\ldots,N-1.
\end{multline}
As an example, an asymmetric 4PSK constellation with $\delta=\pi/3$ is illustrated in Fig. \ref{MPSK_figure}, along with the decision regions for maximum a posteriori (MAP) detecting at the receiver side. For details on the relevant MAP decision rule, see Appendix A.
\par
\begin{figure}
\centering
\includegraphics[width=\linewidth]{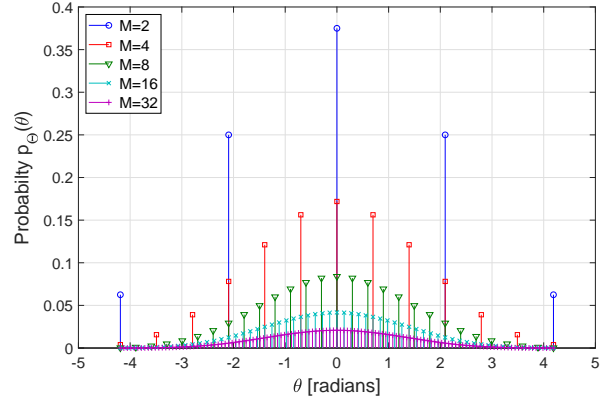}
\caption{The p.m.f. of $\Theta$ for $\delta=\pi/3$}
\label{disc_conv_plot}
\end{figure}

 In analogy to obtaining the p.d.f. of $\Theta$ in \eqref{Theta_eq} 
for the case of continuously distributed phase in Section III.A, the probability mass function (p.m.f.) of $\Theta$ is found for the case of discretely distributed phase by discrete convolution of the p.m.f.s of $\Phi_{n_i}$. The resultant support of $\Theta$ is 
\begin{multline}
\mathrm{supp}(\Theta) = \bigg\{\theta_k=-4\delta+\frac{2\delta k}{M-1},\ \\
k=0,1,\ldots,4(M-1)\ : |\delta|<\pi\bigg\} 
\label{theta1}
\end{multline}
for the case of asymmetric $M$PSK in limited phase range and 
\begin{multline}
\mathrm{supp}(\Theta) = \bigg\{\theta_k=\frac{\left(-4(M-1)+2k\right)\pi}{M},\  \\
k=0,1,\ldots,4(M-1)\ : |\delta|=\pi\bigg\}
\label{theta2}
\end{multline}
for the symmetric $M$PSK. 
As an illustration for the above, Fig. \ref{disc_conv_plot} depicts $p_{\Theta}(\theta)$ for the asymmetric $M$PSK modulation with $\delta=\pi/3$ with uniform input distribution $p_X(x)=1/M$ for different values of $M$.
\par To account for the discrete distribution of the phases $\Phi_{n_i}$, the scaling law in (\ref{Scaling_law_phase}) is modified as follows:
\begin{equation}\label{Scaling_law_phase_discrete}
z_{DC}(\xi) \simeq k_2 R_sP+k_4 R_s^2\frac{(2N^2+1)}{2N}P^2\xi,
\end{equation}
where $\xi=\sum_{k=0}^{4M-4}\cos(\theta_k)\mathrm{Pr}(\Theta=\theta_k)$ 
gives the expected mean of $\cos(\Theta)$ and replaces that in \eqref{cos_theta}, used earlier for the continuous phase distribution.
Note that the received energy is maximized when $\xi=1$.
\par For convenience, the random term in \eqref{Scaling_law_phase_discrete} is rewritten in vector form as  $\xi=\cos(\boldsymbol{\theta})\mathbf{\bar{p}}^T$, where  $\cos(\boldsymbol{\theta})=[\cos(\theta_0)\ \ldots \ \cos(\theta_{4M-4})]$ and $\mathbf{\bar{p}}=[\mathrm{Pr}(\Theta=\theta_0) \ \ldots\ \mathrm{Pr}(\Theta=\theta_{4M-4})]$ are the cosine of the support of the random variable $\Theta$ in \eqref{theta1}-\eqref{theta2} and the corresponding probabilities obtained as $\mathbf{\bar{p}}=\mathbf{p} \ast \mathbf{p}  \ast \mathbf{\acute{p}} \ast \mathbf{\acute{p}}$, respectively. In the latter, $\mathbf{p}=[p_0\ \ldots \ p_{M-1}]$ denotes the p.m.f. of the symbols in $M$PSK constellation  using the notation $p_m=\mathrm{Pr}(X=x_m)$ and $\mathbf{\acute{p}}=[p_{M-1}\ \ldots \ p_0]$ is the 
version of the vector $\mathbf{p}$ with the flipped order of elements. The flipping is necessary to account for the subtractions of the two random variables $\Phi_{n_i}$ in the expression for $\Theta$ in \eqref{Theta_eq}.

\par In comparison to deterministically scaling $z_{DC}(\delta)$ in \eqref{Scaling_law_phase}, the discrete probabilistic nature of $\xi$ in \eqref{Scaling_law_phase_discrete} means that $z_{DC}(\xi)$ not only decreases with increasing the phase range $\delta$, but also depends on the choice of the probability masses of the symbols and may be larger or smaller than in the case of the uniform phase distribution.  It is also noted that for a certain $\delta$, equiprobable modulation of a large enough order $M$ will behave similarly to the uniformly distributed phase, and the scaling laws \eqref{Scaling_law_phase} and \eqref{Scaling_law_phase_discrete} will also nearly match. As a basis for this, consider the p.d.f. and the p.m.f. of $\Theta$ for the same $\delta$ in Figs. \ref{distributions_overlay} and \ref{disc_conv_plot}, respectively. As $M$ 
grows large,  $\ \mathrm{Pr}(\Theta=\theta_k) \rightarrow 0 \ \ \forall k$, which is characteristic of a p.d.f, meaning that the difference in the discrepant fourth order terms in \eqref{Scaling_law_phase} and \eqref{Scaling_law_phase_discrete} fades away. 

\par Given the asymmetric geometry of the constellation as depicted in Fig. \ref{MPSK_figure}, the modulation design of $M$PSK with limited phase range will consist in mainly optimizing the input probability distribution $\mathbf{p}$ for a given $\delta$, determined by the overall objective of achieving certain rate-energy performance. On one hand, from the information-theoretic perspective, this transcribes into choosing asymmetric 
input p.m.f. that allocates higher probability to the outer symbols than to the inner, in order to achieve lower symbol error probability for detecting information. On the other hand, from the energy transfer perspective, the reverse is preferred, with the asymmetric p.m.f. assigning higher probability to the constellation points with a small phase difference, as deduced from Section III.A, with the deterministic single symbol transmission giving the maximum harvested energy. In any case, a non-uniform input distribution is required to balance the rate and energy maximization goals \cite{Zewde_Gursoy}.

\textit{Remark 2:} Asymmetric geometry of constellation appears in the works on coded modulation design for passive backscatter communication in RFID systems \cite{Roy}--\cite{Roy3}. In particular, it is shown that constellation with unequal 
energy per bit in 
in-phase and quadrature components, such as rectangular 4QAM modulation, allows to maximize the harvested power while improving the spectral efficiency of the backscatter uplink, as compared to commonly used binary modulation. 
This highlights that the RF energy harvesting mechanism generally favours asymmetry in modulation throughout a range of WPT applications, which differs from modulation design requirements for wireless information only communication systems.

\textit{Remark 3:} Asymmetry also appeared in the derivation of input distribution for single-carrier SWIPT in \cite{Morteza1,Morteza2}, because asymmetric inputs are characterized by higher order moments.

\subsection{Information Rate Computations} \label{InfoRefSection}
In this subsection, we aim at maximizing the achievable rate when the transmitter utilizes asymmetric $M$PSK modulation.
\subsubsection{Achievable Information Rate}
Consider a multi-carrier transmission with $N$ carriers over a flat fading deterministic channel, i.e. $h_n=1,\ n=0,\ldots,N-1$. We aim at maximizing the mutual information  $\mathrm{I}(\mathbf{X};\mathbf{Y})$ between the input vector $\mathbf{X}=\left[ X_0\ \ldots\ X_{N-1} \right]$ and the output vector $\mathbf{Y}=\left[ Y_0\ \ldots\ Y_{N-1} \right]$, when the transmitter utilizes asymmetric $M$PSK modulation (see equation \eqref{mpsk}) with $s=\sqrt{\gamma}$, where $\gamma$ denotes the signal-to-noise ratio (SNR). We assume that different carriers (subchannels) are statistically independent of each other, i.e. 
$\mathrm{Pr}(\mathbf{Y}|\mathbf{X})=\prod_{n=0}^{N-1}\mathrm{Pr}(Y_n|X_n)$. 
Therefore, under this assumption and due to the fact that $X_n,\ n=0,\ldots,N-1$ are i.i.d., the achievable rate at the receiver is given by
\begin{equation}
I_N \overset{\Delta}{=} \mathrm{I}(\mathbf{X};\mathbf{Y})=N \mathrm{I}(X_0;Y_0).
\label{sum_rate}
\end{equation}
\par Accordingly, in the following, we equivalently focus on maximizing $\mathrm{I}(X_0;Y_0)$. For clarity, in the sequel, we omit the subchannel index. The channel input and output are related through
\begin{equation}
Y=X+Z,
\label{single_channel}
\end{equation}
where $Z$ is an AWGN distributed as $\mathcal{CN}(0,1)$.
\par The capacity of the channel in \eqref{single_channel} is obtained by maximizing the following mutual information
\begin{equation}
I(X;Y)=H(Y)-\log_2(\pi e).
\label{mutual_info}
\end{equation}
The output entropy $H(Y)=-\int \log_2(p_Y(y))p_Y(y) dy$ in \eqref{mutual_info} is computed by using Monte-Carlo or numerical integration \cite{fundamentals}, where the output p.d.f. $p_Y(y)$ is given as
\begin{align}
p_Y(y)&=\sum_{m=0}^{M-1} p_X(x_m)p_Z(y-x_m)\nonumber \\
&= \frac{1}{\pi} \sum_{m=0}^{M-1} p_m e^{ - \left|y -\sqrt{\gamma}e^{j \phi_m}\right|^2}.
\label{p_Y}
\end{align}

\subsubsection{Optimal Input Distribution} 
For $M$PSK with limited phase range, the optimal input distribution $p_X(x)$ needs to be derived in order to calculate the capacity.
\par 
The conditional p.d.f. of the channel output given the channel input for an $M$-ary constellation is
\begin{equation}
p_{Y|X}(y|x_m)=\frac{1}{2\pi\sigma^2}e^{-\frac{|y-x_m|^2}{2\sigma^2}},\  m=0,\ldots,M-1.
\label{Output_Input_pdf}
\end{equation}
Using MAP detector with the detected symbol denoted as $\hat{X}$, the decision region $D_l$ for $\hat{X}=x_l$ for $l=0,\ldots,M-1$ is given by
\begin{equation}
D_l= \{ y=|y|e^{j\phi}: \phi_{L_l} \leq \phi < \phi_{U_l} \},
\end{equation}
where $\phi_{L_l}$ and $\phi_{U_l}$ denote the lower and the upper border phases of the decision region for a given symbol $x_l$.
The transition probabilities for this discrete memoryless channel are obtained as
\begin{align}
\mathrm{Pr}(\hat{X}=x_l|X=x_m)&=\mathrm{Pr}\left(\phi_{L_l} \leq \Phi < \phi_{U_l} |X=x_m \right)\nonumber \\ 
&=\int_{\phi_{L_l}} ^{\phi_{U_l}} p_{\Phi|x_m}(\phi|x_m)d\phi,
\label{transProb}
\end{align}
where the conditional phase p.d.f. is derived from \eqref{Output_Input_pdf} as in \cite{PSKdecision} and is given by
\begin{multline}
p_{\Phi|X}(\phi|x_m)=\frac{1}{2\pi}e^{-\gamma}+\sqrt{\frac{\gamma}{\pi}}\cos(\phi-\phi_m)e^{-\gamma\sin^2(\phi-\phi_m)}\\
\times \left[1-Q\left(\sqrt{2\gamma} \cos(\phi-\phi_m)\right)\right].
\label{phase_pdf}
\end{multline}

For every range of phase, the transition probabilities channel matrix is calculated using \eqref{transProb} and the optimal input distribution can be obtained with the Blahut-Arimoto algorithm \cite{Arimoto}-\cite{Blahut}.

\subsection{Rate-Energy Region Optimization}
\par We can now define the achievable rate-energy region 
(or, more accurately, rate-DC current) region as
\begin{multline}
C_{R-I_{DC}}(p_X(x)) \overset{\Delta}{=} \Bigg\{ (R,I_{DC}): R \leq I_N,  \\
I_{DC}\leq z_{DC}, \sum_{m=0}^{M-1} p_m=1 \Bigg\}.
\end{multline}
\par In order to identify the rate-energy region, one possible way is to formulate the optimization problem as an energy maximization problem subject to rate constraint over the p.m.f. of input symbols, such as
\begin{align}\label{P1}
\max_{\{p_m\}} \hspace{0.3cm} &z_{DC}\\
\textnormal{subject to} \hspace{0.3cm} &I_N \geq R,\\
&\sum_{m=0}^{M-1} p_m=1,
\label{P1_1}
\end{align}
where $z_{DC}$ and $I_N$ are given in \eqref{Scaling_law_phase_discrete} and \eqref{sum_rate}, respectively.  Note that both $z_{DC}$ and $I_N$ are dependent on the phase range $\delta$, number of mass points $M$, values of mass points $\left\{p_m\right\}_{m=0}^{M-1}$ and number of carriers $N$. 

\par The objective function to be minimized is transformed as $f_0=-z_{DC}$ 
 and the rate constraint is set as $f_1=R-I_N \leq 0$. 
The optimization problem \eqref{P1}-\eqref{P1_1} can then be rewritten as 
\begin{align}\label{P2}
\min_{\mathbf{p}} \hspace{0.3cm} &f_0(\mathbf{p})\\
\textnormal{subject to} \hspace{0.3cm} &f_1(\mathbf{p})\leq 0,\\
&\mathbf{1}^T\mathbf{p}-1=0,
\label{P2_end}
\end{align}
where $\mathbf{p}$ is the vector representation of $\left\{p_m\right\}_{m=0}^{M-1}$.
\par The rate constraint is convex in $\mathbf{p}$ due to the concavity of the output entropy $H(Y)$ when $I_N$ is evaluated numerically over a restricted domain. However, the objective function is non-convex, as it is in essence a signomial function in the variables 
$\{p_m\}$ as some coefficients $\cos(\theta_k)$ of the products of powers of $\{p_m\}$ are negative in $\xi$ in \eqref{Scaling_law_phase_discrete} (see Appendix B for an example). Because the rate constraint cannot be represented as a posynomial or a signomial, the obtained optimization problem \eqref{P2}-\eqref{P2_end} is not compatible with standard Signomial Geometric Programming (SGP) tools \cite{GP_Chiang}. 
However, a locally optimal solution can still be obtained efficiently with Sequantial Quadratic Programming (SQP) algorithm \cite{SQP} that is a quasi-Newton method for solving inequality-constrained nonlinear programming problems. For details on the gradients of the energy and rate functions as supplied in SQP optimization, see Appendix C.

\section{Simulation Results}\label{section_simulations}
We first present analytical and numerical results for harvested energy with multi-carrier transmission and uniformly distributed random phases. 
Next, focussing on the aspect of WIT only, we obtain the optimal input distribution for the proposed asymmetric $M$PSK modulation and calculate the maximum achievable information rate. We then show optimization results for the rate-energy region with asymmetric $M$PSK modulation. 
Finally, we analyse the simulation results for the symbols' p.m.f. optimized for SWIPT and how harvested energy scales with the modulation order $M$.
\begin{figure}
\centering
\includegraphics[width=0.9\linewidth]{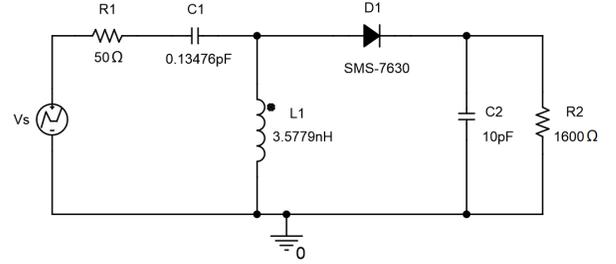}
  \caption{Rectenna with a single series diode rectifier.}
  \label{rectenna_circuit}
\end{figure}
\par All SWIPT observations are confirmed by PSpice simulations of a practical rectenna circuit. The rectenna design (Fig. \ref{rectenna_circuit}) is optimized for the input signal composed of 4 in-phase carriers centered around 5.18GHz and the available RF power $P_{in,av}$ of $-20$dBm. The package parasitics of components are ignored. 
The L-matching network is optimized together with the load resistor, with the objective to maximize the output DC power and minimize impedance mismatch due to a signal of varying instantaneous power.

\subsection{Results for Multi-Carrier Transmission with Uniformly Distributed Random Phase}

\begin{figure}
\centering
\includegraphics[width=\linewidth]{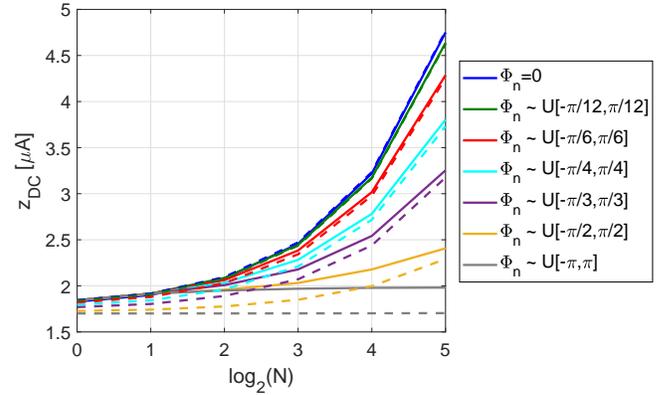}
\caption{$z_{DC}$ obtained numerically (solid lines) and with the scaling law (dash lines) as a function of $\log_2(N)$.}
\label{zdc_analytical_and_simulated}
\end{figure}
\par Fig. \ref{zdc_analytical_and_simulated} represents $z_{DC}$ as a function of the number of carriers $N$ with the phase of each carrier chosen to be i.i.d. distributed as $\Phi_{n} \sim U[-\delta,\delta]$. The value of $z_{DC}$ is obtained numerically by averaging over several hundred symbol periods using \eqref{zdc}. The 
power $P$ is set as $-20$dBm and $N$ carriers are centered around 5.18GHz with a frequency gap fixed as $\Delta_f=B/N$ for bandwidth $B=10$MHz. The symbol period is set as $T=1/ \Delta f$. Fig. \ref{zdc_analytical_and_simulated} also illustrates the scaling law in \eqref{Scaling_law_phase} for $z_{DC}$ as a function of the number of carriers $N$ for different values of $\delta$. As it is observed, apart from the small $N$ region, there is a good match between the analytical and the numerical results. This inaccuracy for small values of $N$ can be justified by the following fact. When the carriers with the same index $n_i=n_j$ (and, consequently, with the same phase $\Phi_{n_i}=\Phi_{n_j}$) are combined in the summation in \eqref{4thOrderSumVaryingPhase}, they contribute constructively to the channel output. This effect is particularly pronounced for small values of $N$ and is not captured by the analytical model in \eqref{Scaling_law_phase}. The small variations in the analytical and the numerical $z_{DC}$ for $\delta=\pi$ are due to the same aforementioned reason.

\begin{figure}
\centering
\includegraphics[width=\linewidth]{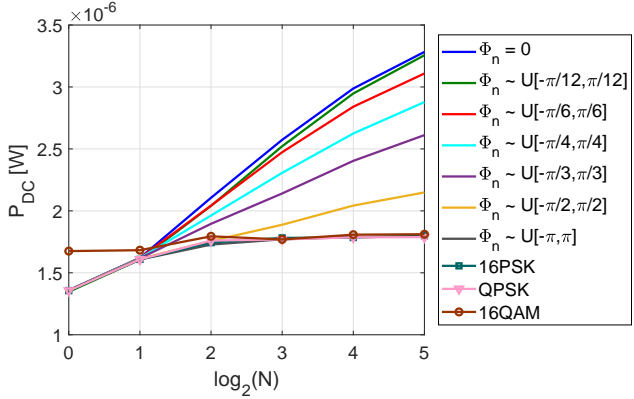}
\caption{Average DC power $P_{DC}$ as a function of $\log_2(N)$ for input signals with  with i.i.d. uniformly distributed phases.}
\label{Pdc_pspice_for_varying_phase}
\end{figure}
\par To validate the scaling laws of Fig. \ref{zdc_analytical_and_simulated}, the rectenna circuit of Fig. \ref{rectenna_circuit} is simulated by using the 
input data signals with uniformly distributed random phase. The PSpice simulations results in terms of the DC output power (Fig. \ref{Pdc_pspice_for_varying_phase}) confirm the analytical and the numerical results using the $z_{DC}$ metric (Fig. \ref{zdc_analytical_and_simulated}). 
The $P_{DC}$ behaviour naturally has a more saturated form due to the non-optimality of the circuit design for large $N$ (because of the choice of the finite output capacitor and load).

\par
It is verified that, in the nonlinear operating region of the diode at low input power, the circuit is very sensitive to randomness in phase
, with the losses due to phase variations increasing with $N$. On the other hand, amplitude randomness can be beneficial for energy harvesting purposes at low input power, due to high power waveform peaks driving the diode with higher efficiency. As a result, it is observed that 16QAM is slightly better than PSK for small $N$ in accordance with \cite{Mod_Scheme_RF_DC_Japan}, but as $N$ grows, there are less high power peaks 
for QAM and this benefit of 16QAM over PSK is lost.
In addition, it is noted that QPSK and 16PSK modulations with equiprobable signaling perform equivalently to the case of $\Phi_{n} \sim U[-\pi,\pi]$.  This highlights that the constellation size of symmetric $M$PSK modulation with uniform p.m.f. has no effect on harvested energy, because in such case $\xi=0$ in the fourth order component of \eqref{Scaling_law_phase_discrete} resulting in $z_{DC} \simeq k_2R_sP$, same as for $\Phi_n \sim U [-\pi, \pi]$. 

\subsection{Results for Information Rate with Asymmetric $M$PSK Modulation}
\par
\begin{figure}
\centering
\includegraphics[width=\linewidth]{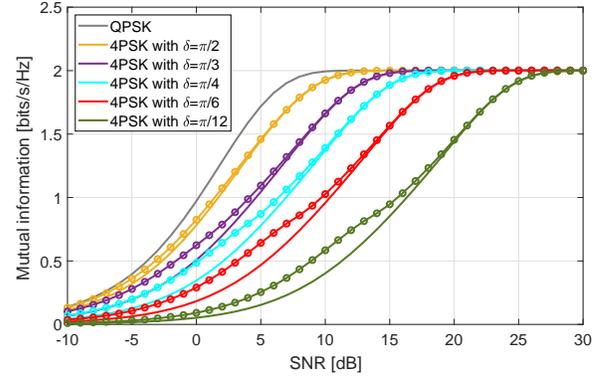}
\caption{Mutual information for 4PSK in different ranges $[-\delta,\delta]$ with uniform and optimal input distributions. (Marked lines correspond to $p_X(x)$ optimized with the Blahut-Arimoto algorithm).}
\label{capacityM4}
\end{figure}

 Considering a single WIT channel, in Fig. \ref{capacityM4}, the mutual information is obtained using \eqref{mutual_info} for different ranges $[-\delta,\delta]$ of 4PSK constellation with uniform and optimal input distributions, where the latter is computed using the Blahut-Arimoto algorithm \cite{Arimoto}--\cite{Blahut} and the channel transition probabilities given by \eqref{transProb}. It is noted that for asymmetric $M$PSK modulation, the equiprobable input distribution $p_X(x)$ is not optimal anymore and optimizing $p_X(x)$ with the Blahuto-Arimoto algorithm further maximizes the mutual information.  As an example, for $\delta=\pi/4$ at $\gamma=10$dB, the outer symbols have been optimized to probabilities $p_0=p_3=0.327$ and the inner symbols -- to probabilities $p_1=p_2=0.173$. It is also noted that the input distribution converges to the uniform distribution at high SNR values. Finally, it is observed that such choice of $\delta$ can be made for a given SNR, which does not lead to a large gap between the maximum information rate of the symmetric $M$PSK modulation and that achievable with the asymmetric $M$PSK modulation. Moreover, at high SNR values, the rate close to that of the symmetric $M$PSK modulation is achievable with even small $[-\delta,\delta]$ ranges.

\subsection{Results for Rate-Energy Region with Asymmetric $M$PSK Modulation}
\begin{figure*}
\centering
\includegraphics[width=0.8\linewidth]{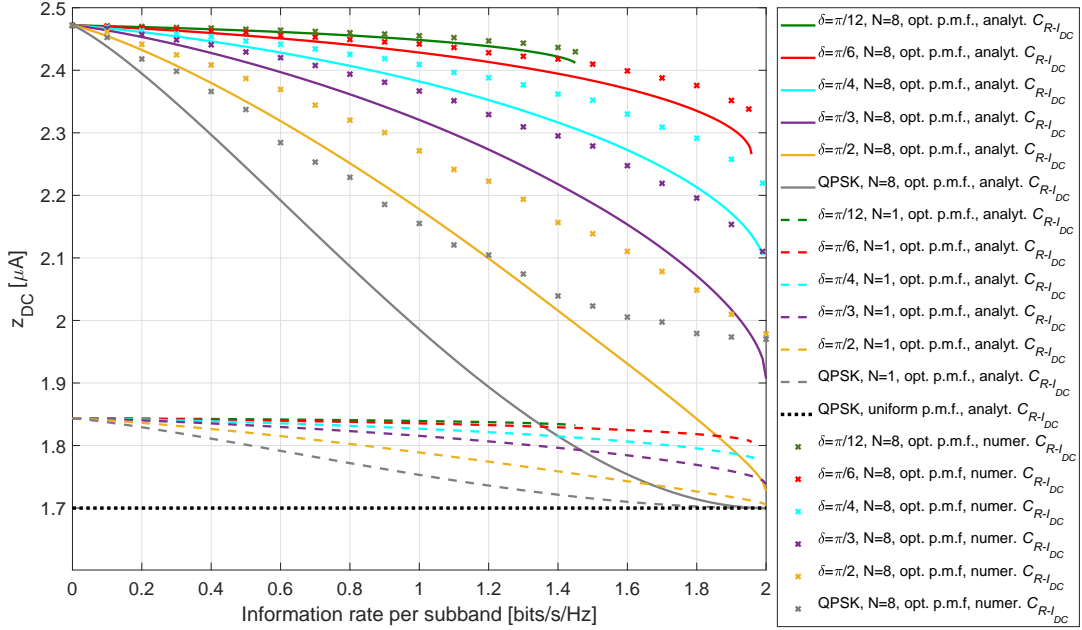}
\caption{Analytical $C_{R-I_{DC}}$ for $N=$\{$1,8$\} and numerical $C_{R-I_{DC}}$ for $N=8$ as a function of $\delta$ for asymmetric 4PSK modulation with optimized (opt.) p.m.f. The lower bound for analytical $C_{R-I_{DC}}$ is given by QPSK with uniform p.m.f.}
\label{zdc_r_n1_8}
\end{figure*}
The rate-energy region is obtained 
by solving the optimization problem (\ref{P1})-(\ref{P1_1}) with the help of MATLAB Optimization Toolbox and the choice of SQP for \textit{fmincon} algorithm. 
\par Fig. \ref{zdc_r_n1_8} illustrates the achievable rate-energy region for 4PSK in various phase range limits for a single ($N=1$) and multiple ($N=8$) carrier transmission and 
SNR of $20$dB. The total information rate $I_N$ is normalized w.r.t. the bandwidth $NB_s$. Hence, the x-axis refers to a per-subband rate. The gains due to using optimized modulation and multi-carrier waveform contribute both to the increase in $z_{DC}$. Transmitting over multiple carriers 
results in higher ECEs -- this benefit 
is added onto the gains due to optimization of the symbols' p.m.f. relative to the non-optimized QPSK with equiprobable signaling. 
As such, QPSK with uniform p.m.f. provides a lower bound for harvested energy, which is constant for all values of rate, irrespective of the number of carriers. In general, symmetric $M$PSK with uniform p.m.f. gives $\xi=0$, meaning that there is no fourth order contribution in $z_{DC}$ in \eqref{Scaling_law_phase_discrete} and the same energy-rate per subband performance is predicted analytically for any $N$ (and $M$ as discussed in Section IV.A). In practice, this is true for any large $N$ (e.g. $N\geq 8$ as observed from Fig. \ref{Pdc_pspice_for_varying_phase}) due to random modulations of phase at each carrier resulting in a random aperiodic time-domain waveform with low instantaneous PAPR that is unable to deliver high ECE at the rectifier \cite{Clerckx:SWIPT_long}. 

\begin{figure}
\centering
\includegraphics[width=\linewidth]{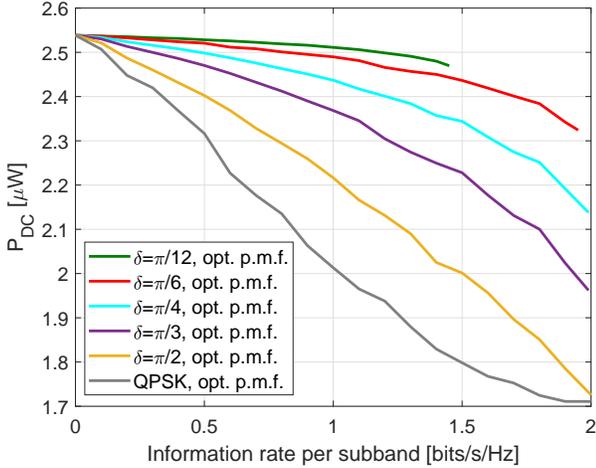}
\caption{$R-P_{DC}$ region as a function of $\delta$ for $N=8$ and $M=4$.}
\label{R_PDC}
\end{figure}

\par It is also noted that it is possible to significantly enlarge the rate-energy region when the phase range of symbols is limited, i.e. $M$PSK constellation is concentrated in only a fraction of the circle on the complex plane.  It is observed that the rate-energy region is maximized most with $\delta=\pi/6$, giving the best value of $\delta$ for this setup out of the tested values. It can also be derived analytically that for all values of $\delta$, $\xi=1$ when only one symbol is transmitted with probability 1 giving the maximum energy point of $z_{DC}=2.47 \mu A$, with the full details on the resultant p.m.f. given further in Section IV.\ref{section_mpf} In accordance with Fig. \ref{capacityM4}, smaller phase ranges with $\delta=\{\pi/12,\pi/6\}$ do not allow to achieve the maximum normalized rate of 2 bits/s/Hz at the SNR of 20dB. The resultant optimized p.m.f. of symbols also converges to that obtained with the Blahut-Arimoto algorithm at maximum rate points. 

\par The simulation-based rate-energy region is also shown in Fig. \ref{zdc_r_n1_8} for $N=8$, verifying the suitability of the EH $z_{DC}$ model \eqref{Scaling_law_phase_discrete} for performing optimization of asymmetric 
$M$PSK modulation. In order to obtain the 
numerical results, $z_{DC}$ was calculated as the time average using \eqref{zdc}. The averaging is performed over several hundred symbol periods, with symbols transmitted independently on each carrier and the symbol probabilities optimized for the tested phase ranges and at a given rate requirement. As $N=8$ used in the comparison of the analytical and the numerical rate-energy regions is rather small, a gap between the regions' boundaries of the order similar to the one in Fig. \ref{zdc_analytical_and_simulated} is observed. This inaccuracy occurs due to the reason mentioned in Section IV.A, whereby for small $N$, the constructive combining of the carriers with the same phase is not captured by the analytical model \eqref{Scaling_law_phase_discrete}. Nevertheless, this does not affect the optimization results, as essentially the objective is to maximize the component $\xi$, which is independent of $N$, in $z_{DC}$ in \eqref{Scaling_law_phase_discrete}.

\begin{figure}
\centering
\includegraphics[width=1\linewidth]{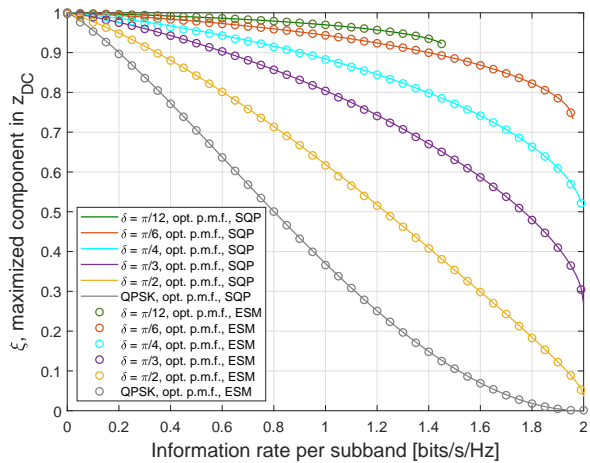}
\caption{$R-\xi$ region as a function of $\delta$ for $M=4$, with the optimal p.m.f. obtained with SQP and ESM algorithms.}
\label{SQP_ESM}
\end{figure}

\par In order to validate the results in Fig. \ref{zdc_r_n1_8}, the single series rectifier circuit (Fig. \ref{rectenna_circuit}) is simulated by reusing the input data signals generated for $N=8$, with the p.m.f. of 4PSK symbols optimized for a set of $\delta$ values. 
The rate-$P_{DC}$ region graph in Fig. \ref{R_PDC} confirms the analytical and the numerical rate-$z_{DC}$ region results in Fig. \ref{zdc_r_n1_8}. This validates the EH model and modulation design and highlights the importance of using optimized modulation design in order to increase the ECE of a practical rectenna circuit in the 
SWIPT configuration.

\par Finally, the results of the optimization using SQP are compared with the globally optimal solution obtained with the exhaustive search method (ESM) in Fig. \ref{SQP_ESM}. The rate-energy regions are presented as normalized rate-$\xi$ regions, where $\xi$ is the component in \eqref{Scaling_law_phase_discrete} that needs to be maximized for energy transfer.  It is noted that the two sets of results match well, verifying the suitability of the SQP algorithm with an appropriate choice of the initialization point to solve the optimization problem \eqref{P1}-\eqref{P1_1} efficiently. 

\subsection{Analysis of Optimized Symbols P.M.F.}\label{section_mpf}

\begin{figure*}
\centering
\includegraphics[width=0.75\linewidth]{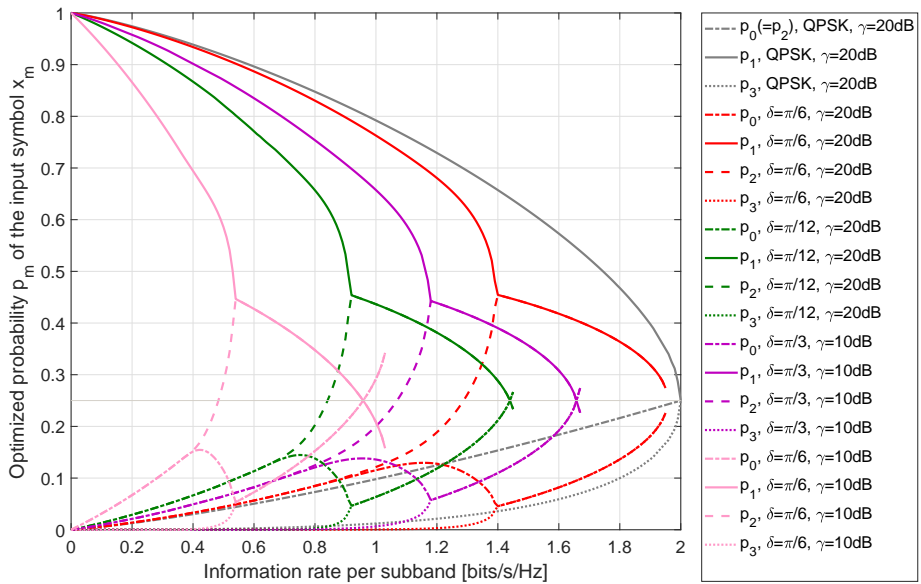}
\caption{Optimized probabilities of 4PSK input symbols for $\gamma=\{20,10\}$dB and various $\delta$ as a function of information rate.}
\label{po_vs_pi}
\end{figure*}

\par Fig. \ref{po_vs_pi} shows the resultant p.m.f. of the 4PSK symbols for a set of $\delta$ and SNR values. The plot displays locally optimal optimization results. Deriving conditions under which the maximum energy is attainable eliminates the need for the random start during optimization. As $\xi=1$ is achievable when only one symbol is transmitted, choosing the initial start p.m.f. as, for instance, $[0\ 1\ 0\ 0]$ allows to obtain the rate-energy region boundaries efficiently.

\par The maximum gain in terms of energy 
is achievable in the lower rate region when one symbol (e.g. $x_1$) is allocated dominant probability, with the two closest symbols ($x_0$ and $x_2$) getting equal probability ($p_0=p_2$) to gradually increase the rate, which follows from the expression for $\Theta$ \eqref{Theta_eq}. Equivalently, other subset of neighbouring symbols could provide the same result. The phase shifts due to the two symbols near the dominant one compensate each other on average. As the rate requirement rises, the full constellation becomes utilized and the optimization pattern changes for the case of the asymmetric 4PSK. The two inner symbols ($x_1$ and $x_2$) and the two outer symbols ($x_0$ and $x_3$) now get equal probability, i.e. $p_1=p_2$ and $p_0=p_3$, and the corresponding plots of the probabilities coincide. For the standard QPSK constellation, the pattern does not change and $p_0=p_2$ applies for the full range of rate.

\par For the case of the asymmetric PSK, as the requirement on rate increases and the requirement on energy decreases, the optimization results in the reallocation of probability weights from the inner symbols maximizing energy -- to the outer symbols maximizing rate, in accordance with the analysis in Sections III.B and III.C and the simulation results in Section IV.B. At a relatively high energy point, the inner symbols get allocated most of the probability mass, $\approx 0.45$ each. Depending on the SNR and $\delta$, the outer symbols gain more weight as the required rate increases, until the p.m.f. becomes near uniform ($p_m\approx 0.25$) at the maximum rate point at 20dB SNR and at a lower rate point at 10dB SNR. For the higher SNR value, the p.m.f. of the input symbols is close to uniform for all $\delta$ values, when the rate is at maximum. For the lower SNR value, 
the demand on rate results in the probability plot of the outer symbols overshooting the probability plot of the inner symbols, with the overshoot increasing and occurring for lower rate as $\delta$ decreases.

The underlying physical process behind the energy maximization is that when carriers carry symbols with relatively small phase differences, the carrier peaks are closer in phase at the receiver giving a larger peak amplitude of the multi-carrier waveform, which in turn is able to produce a higher voltage at the output of the diode-based rectifier, due to its non-linear behaviour. However, with the active rate constraint, this mechanism needs to be balanced with improving the error probability of symbol detection, thereby gradually allocating the outer symbols more probability mass as the rate is increased.

\subsection{Analysis of Rate-Energy Scaling with Modulation Order $M$}
\begin{figure}
\centering
\includegraphics[width=\linewidth]
{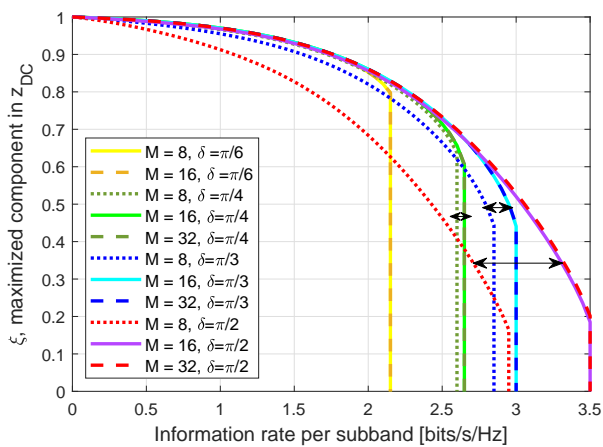}
\caption{$R-\xi$ region as a function of $M$ and $\delta$, $\gamma=20$dB. 
The black arrows indicate the gaps between the boundaries of the $R-\xi$ regions for a given $\delta$ that can be overcome by increasing $M$.}
\label{R_E_region_M}
\end{figure}

\begin{figure}
\centering
\includegraphics[width=\linewidth]{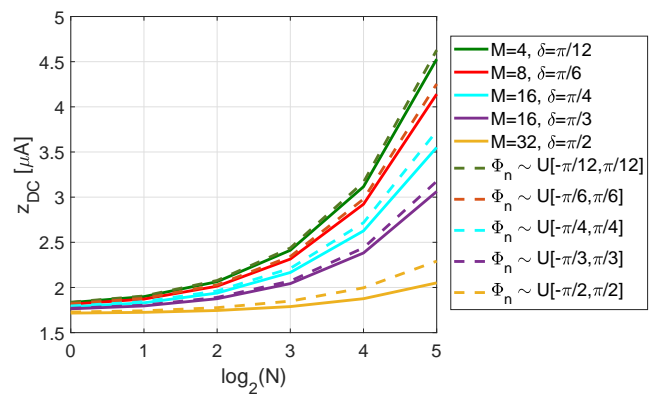}
\caption{$z_{DC}$ (solid lines) using asymmetric $M$PSK modulation, optimized for minimum energy -- maximum rate points, $\gamma=20$dB, and $z_{DC}$ (dash lines) obtained with the scaling law for the uniformly distributed phase, as a function of $\log_2(N)$.}
\label{E_of_M_scaling}
\end{figure}
\par Finally, for every $\delta$ the rate-energy region can only be enlarged up to a certain value of $M$, and increasing $M$ further would not bring any significant benefit at a given SNR. As shown in Fig. \ref{R_E_region_M}, the boundaries of the largest regions achievable for every $\delta$ can, for example, be reached with conditions  
$(\delta=\pi/6$; $M=8)$, $(\delta=\pi/4$; $M=16)$, $(\delta=\pi/3$; $M=16)$, $(\delta=\pi/2$; $M=32)$. 
\par Fig. \ref{E_of_M_scaling} shows the scaling of $z_{DC}$ with $\log_2(N)$ calculated at minimum energy -- maximum rate points,
when optimized input symbol distributions turn out to be close to uniform at the SNR of 20dB. It is observed that $z_{DC}$ obtained with the asymmetric $M$PSK modulation 
follows the scaling law (\ref{Scaling_law_phase}) derived for $\Phi_{n} \sim U[-\delta,\delta]$, illustrating the analysis in Section III.B, and Fig. \ref{E_of_M_scaling} can be related to the analytical results in Fig. \ref{zdc_analytical_and_simulated}, replotted here (in Fig. \ref{E_of_M_scaling}) for convenience of comparison. It is noted that $z_{DC}$ obtained with the scaling law \eqref{Scaling_law_phase} for the uniformly distributed phase is slightly higher than $z_{DC}$ with the asymmetric $M$PSK modulation. This difference in plots is attributed to the fact that the optimized symbol probabilities are close to but not exactly uniform
, as observed in Fig. \ref{po_vs_pi} and discussed in Section IV.D. It is demonstrated that the scaling law (\ref{Scaling_law_phase}) matches the scaling law \eqref{Scaling_law_phase_discrete} in case of the uniform p.m.f. of $M$PSK symbols (that is optimal for different rate-energy points at different SNRs), provided that $M$ is above a certain limit $\bar{M}$ for a given $\delta$. In such case, at high SNR values, the scaling law (\ref{Scaling_law_phase}) approximates a lower bound for $z_{DC}$ using asymmetric $M$PSK modulation for SWIPT.

To summarize, the choice of different parameters of the asymmetric $M$PSK modulation has an effect on the rate and energy delivered to the SWIPT system using multi-carrier transmission: larger phase range $\delta$ increases the information rate but decreases the harvested energy; greater modulation order $M$ up to a limit $\bar{M}$ enlarges the whole rate-energy region; and increasing the number of carriers scales up the energy for the same information rate per subband.

\section{Conclusions}\label{section_conclusion}
In this paper, we proposed the asymmetric PSK modulation, specifically designed for SWIPT. 
We optimized the probabilities of symbols for the symmetric and the presented asymmetric PSK modulations, thereby enlarging the SWIPT rate-energy region considerably as compared to the equiprobable symmetric PSK.
Such modulation design is enabled by using the accurate nonlinear model for the rectifier at the energy harvesting and information decoding receiver. A unified information- and energy-carrying multi-carrier waveform is used for the purpose of SWIPT. The gain in the harvested energy due to using optimized modulation design is additive to the increase in the ECE achieved by utililizing multiple carriers. 
\par The trade-off between harvested energy and information rate manifests itself in the optimization of the 
input probabilities of the PSK symbols, which are different for the vertices of the rate-energy region, depending on whether maximum rate or energy is achieved, and change in-between. This is due to the conflicting mechanism inherent to SWIPT, whereby WPT requires phases of carriers to be equal, i.e. transmitting only one symbol, whereas WIT aims at maximizing the rate by using the constellation in full, while minimizing the probability of error between symbols. 
\par An interesting extension of this work would be adapting the asymmetric PSK modulation for the frequency-selective channel. The nonlinear EH model could also be applied in optimizing other practical modulation schemes for SWIPT, by examining not only phase but also amplitude asymmetry of the constellation. Among other open problems for future research, is deriving practical modulation design for power splitting and time switching SWIPT receivers, examining the suitability of asymmetric PSK for backscatter communication, as well as designing error-correcting codes for the new modulation schemes.

\appendices
\section{MAP Detector for Asymmetric $M$PSK Modulation}
Denoting the detected output symbol as $\hat{X}$ and omitting the subchannel index, we derive a decision rule for the MAP detector at the receiver as 
\begin{equation}
\hat{X}(y)=\argmax_{\left\{x_m\right\}_{m=0}^{M-1}}  p_{Y|X}(y|x_m) \mathrm{Pr}(X=x_m),
\end{equation}
or equivalently, in terms of phase for the asymmetric $M$PSK modulation,
\begin{equation}
\hat{X}(\phi)=\argmax_{\left\{x_m\right\}_{m=0}^{M-1}}  p_{\Phi|X}(\phi|x_m) \mathrm{Pr}(X=x_m),
\end{equation}
where $p_{\Phi|X}(\phi|x_m)$ is obtained as in \eqref{phase_pdf}.
%

\section{Example of Signomial Term in $z_{DC}$ for $M$PSK}
Consider an example of $M$PSK modulation with limited phase range and $M=4$. By calculating 
the p.m.f. $\mathbf{\bar{p}}=\mathbf{\tilde{p}} \ast \mathbf{\tilde{p}}$ with $\mathbf{\tilde{p}} \overset{\Delta}{=} \mathbf{p} \ast \mathbf{\acute{p}}$, corresponding to the random variable $\Theta$, we have
\begin{multline}
\mathbf{\tilde{p}}=[p_0p_3,\ (p_0p_2+p_1p_3),\ (p_0p_1+p_1p_2+p_2p_3),\ (p_0^2+p_1^2\\ +p_2^2+p_3^2),\ (p_0p_1+p_1p_2+p_2p_3),\ (p_0p_2+p_1p_3),\ p_0p_3]
\end{multline}
and
\begin{multline}
\mathbf{\bar{p}}=[\tilde{p_0}^2,\ 2\tilde{p_0}\tilde{p_1},\ (2\tilde{p_0}\tilde{p_2}+\tilde{p_1}^2),\ 2(\tilde{p_0}\tilde{p_3}+\tilde{p_1}\tilde{p_2}),\ 
(2\tilde{p_0}\tilde{p_4}\\+2\tilde{p_1}\tilde{p_3}+\tilde{p_2}^2),\ 2(\tilde{p_0}\tilde{p_5}+\tilde{p_1}\tilde{p_4}+\tilde{p_2}\tilde{p_3}),\ 
(2\tilde{p_0}\tilde{p_6}+2\tilde{p_1}\tilde{p_5}\\+2\tilde{p_2}\tilde{p_4}+\tilde{p_3}^2),\ 2(\tilde{p_1}\tilde{p_6}+\tilde{p_2}\tilde{p_5}+\tilde{p_3}\tilde{p_4}),\ 
(2\tilde{p_2}\tilde{p_6}+2\tilde{p_3}\tilde{p_5}\\+\tilde{p_4}^2),\ 2(\tilde{p_3}\tilde{p_6}+\tilde{p_4}\tilde{p_5}),\ (2\tilde{p_4}\tilde{p_6}+\tilde{p_5}^2),\ 2\tilde{p_5}\tilde{p_6},\ \tilde{p_6}^2].
\end{multline}
The resultant $\xi=\cos(\boldsymbol{\theta})\mathbf{\bar{p}}^T$ is in general a signomial in $p_m$, since some coefficients $\cos(\theta_k) < 0$. For a particular case of $M=4$, $\xi$ is a posynomial for a small range of phase ($\delta \lessapprox \pi/8$).

\section{Gradients of the Energy and Rate Functions}
\par The details of the analytical expressions for gradients $\nabla f_0(\mathbf{p})$ and $\nabla f_1(\mathbf{p})$, as supplied in SQP optimization, are given below.

\par In order to compute the gradient of the energy function $\nabla f_0(\mathbf{p})$, the partial derivatives $\frac{\partial \xi}{\partial p_m}$ 
are obtained as 
\begin{equation}
\frac{\partial \xi}{\partial p_m}=\cos(\boldsymbol{\theta})\frac{\partial \bar{\mathbf{p}}^T}{\partial{p_m}}.
\end{equation}
\par The implementation of the gradient of the rate constraint function $\nabla f_1(\mathbf{p})$ 
relies on taking a double integral over the real and imaginary parts of the received symbol $y$ in order to compute $H(Y)$ numerically. The output entropy $H(Y)$ in \eqref{mutual_info} is given as
\begin{multline}
H(Y)=-\int_{-\infty}^{\infty}\int_{-\infty}^{\infty}\log_2\Bigg(\sum_{m=0}^{M-1}\frac{p_m}{\pi} \exp \bigg(- \Big( y_{re}\\
-\sqrt{\gamma} \cos (\phi_m)\Big)^2-\Big(y_{im}-\sqrt{\gamma}\sin (\phi_m)\Big)^2\bigg)\Bigg) \sum_{m=0}^{M-1}\frac{p_m}{\pi} \exp \bigg(\\
- \Big( y_{re}-\sqrt{\gamma} \cos (\phi_m)\Big)^2-\Big(y_{im}-\sqrt{\gamma}\sin (\phi_m)\Big)^2\bigg)dy_{re}dy_{im}
\end{multline}
and using the theorem of Dominated Convergence, the partial derivatives $\frac{\partial H}{\partial p_m}$ are obtained as
\begin{multline}
\frac{\partial H}{\partial p_m}=-\int_{-\infty}^{\infty}\int_{-\infty}^{\infty}\frac{1}{\pi}\exp \bigg(- \Big( y_{re}-\sqrt{\gamma} \cos (\phi_m)\Big)^2-\Big(y_{im}\\
-\sqrt{\gamma}\sin (\phi_m)\Big)^2\bigg)\Bigg[\frac{1}{\ln 2}+
\log_2\left(\sum_{m=0}^{M-1}\frac{p_m}{\pi} \exp \bigg(- \Big( y_{re} \right.\\
-\sqrt{\gamma} \cos (\phi_m)\Big)^2-\Big(y_{im}-\sqrt{\gamma}\sin (\phi_m)\Big)^2 \bigg) \Bigg) \Bigg]dy_{re}dy_{im}.
\end{multline}

\end{document}